\documentclass{article}
\usepackage{spconf,amsmath,amssymb,graphicx}
\usepackage{multirow}
\usepackage{bbm}
\usepackage{booktabs}
\usepackage{float}
\usepackage{hyperref}
\usepackage{multicol, blindtext}
\usepackage{enumitem}
\usepackage{comment}
\usepackage{graphicx}
\usepackage{xcolor}
\usepackage[dvipsnames]{xcolor}

\def\y{\ensuremath{\mathbf{y}}}

\title{BUT System Description for CHiME-9 MCoRec Challenge}

\name{Dominik Klement, Alexander Polok, Nguyen Hai Phong, Prachi Singh, Lukáš Burget}
\address{Speech@FIT, Brno University of Technology, Czech Republic}

\begin{document}
\ninept 

\maketitle

\begin{abstract}
Multi-talker automatic speech recognition (ASR) in conversational recordings remains an open problem, particularly in scenarios with large portion of overlapping speech where identifying and transcribing a target speaker is difficult from audio alone. Visual cues can help resolve speaker ambiguity, yet their integration into long-context audio-visual (AV) ASR systems has been limited. The CHiME-9 MCoRec task addresses this challenge by requiring transcription of audio-visual recordings of heavily-overlapped parallel conversations, followed by clustering the participants into conversational groups.
In this work, we present the BUT system based on a long-context target-speaker AV-ASR model capable of processing long-form recordings in a single decoding pass. Our architecture conditions a pre-trained NVIDIA Parakeet-v2 ASR model on visual representations from a pre-trained AV-HuBERT model. To cluster participants into conversation groups, we employ Qwen3.5-122B LLM to estimate transcript topic similarity followed by hierarchical agglomerative clustering.
On the MCoRec development set, the proposed system achieves 33.7\% WER and a clustering F1 score of 0.97, improving over the official baseline by 16.2\% WER and 0.15 F1 absolute. 
On the eval set, our team ranked second, being 0.16\% WER and 0.5\% F1 worse than the best system.\footnote{\href{https://www.chimechallenge.org/current/task1/results}{https://www.chimechallenge.org/current/task1/results}}
\end{abstract}

\begin{keywords}
CHiME challenge, Audio-visual speech recognition, target-speaker speech recognition, LLM-driven clustering, multi-modal speech processing
\end{keywords}

\section{Introduction}
\label{sec:pipeline_overview}
Transcription of multi-party conversations is a fundamental problem in speech processing, complicated by overlapping speech, rapid turn-taking, and reverberant environments~\cite{watanabe2020chime, cornell2023chime, vinnikov24_interspeech}. The CHiME-9 MCoRec task~\cite{nguyen2025cocktail} presents a particularly challenging instance of this problem: multiple simultaneous conversations occur within the same recording, creating a "cocktail party" scenario challenging for human listeners and systems submitted to previous CHiME challenges~\cite{polok2024chime8, polok2025dicow, niu24_chime,ntt_chime8,stcon_chime8}. The task requires participants not only to provide speaker-attributed transcriptions in these adverse acoustic conditions but also to correctly cluster speakers into their respective conversation groups.

Audio-only systems can be improved by incorporating additional modalities, such as visual cues from lip movements, which provide a complementary, noise-robust signal enhancing speaker separation and recognition in acoustically challenging environments.
Recent progress in this area has been significant, with state-of-the-art audio-visual (AV) ASR systems~\cite{auto-avsr, rouditchenko24_interspeech} achieving very low Word Error Rates (WER) of 1.5\% and 0.9\% on standard benchmarks LRS2~\cite{LRS2} and LRS3~\cite{afouras2018lrs3}, respectively.
However, these models often fail when applied to realistic, uncurated scenarios characterized by natural noise and unconstrained conversation dynamics~\cite{nguyen25b_interspeech}.

In this paper, we describe the BUT submission to the CHiME-9 MCoRec task.
Our contributions are threefold. First, we propose a long-form audio-visual target-speaker ASR (AV-TS-ASR) system that conditions a pre-trained Nvidia Parakeet model~\cite{sekoyan2025canary1bv2parakeettdt06bv3efficient} on visual features from AV-HuBERT~\cite{avhubert}. Second, we introduce conversation clustering with an LLM-driven semantic approach. Third, we demonstrate substantial improvements over the official baseline in both transcription and clustering performance on CHiME-9 MCoRec.
Our pipeline is available on GitHub.\footnote{\href{https://github.com/BUTSpeechFIT/CHiME-9-AV-TS-ASR}{https://github.com/BUTSpeechFIT/CHiME-9-AV-TS-ASR}}

\section{Proposed System}
\label{sec:method}

\begin{figure}[htb]
    \centering
    \includegraphics[width=1\linewidth]{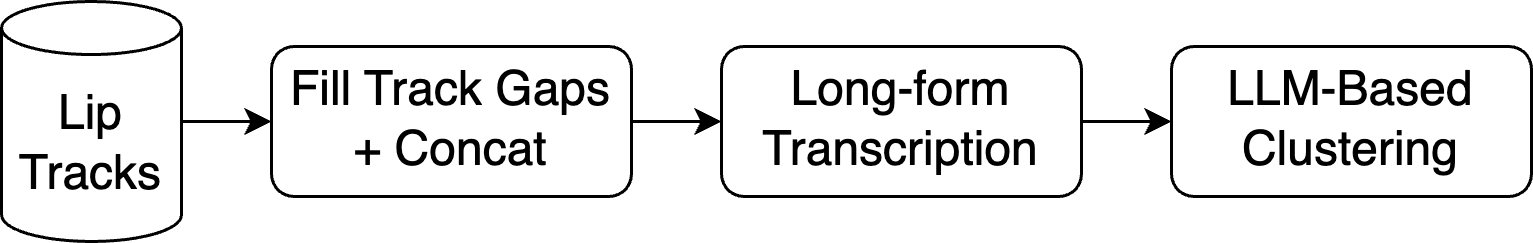}
    \caption{Overview of the proposed pipeline.}
    \label{fig:overview_diag}
\end{figure}

Our system, illustrated in Figure~\ref{fig:overview_diag}, follows the general pipeline of the official challenge baseline~\cite{nguyen25b_interspeech, nguyen2025cocktail}, with modifications to both transcription and conversation-group clustering. To enable long-form processing, we fill missing face-detection frames with black frames and decode the concatenated audio-visual input in a single pass. Conversation groups are then inferred using a clustering pipeline based on topic similarity estimated by the Qwen3.5 large language model (LLM) from the target-speaker transcripts~\cite{yang2025qwen3}.

\vspace{-10pt}

\subsection{Audio-Visual Target-Speaker ASR}
\label{sec:av_ts_asr}

\begin{figure}[t]
    \centering
    \includegraphics[width=1\linewidth]{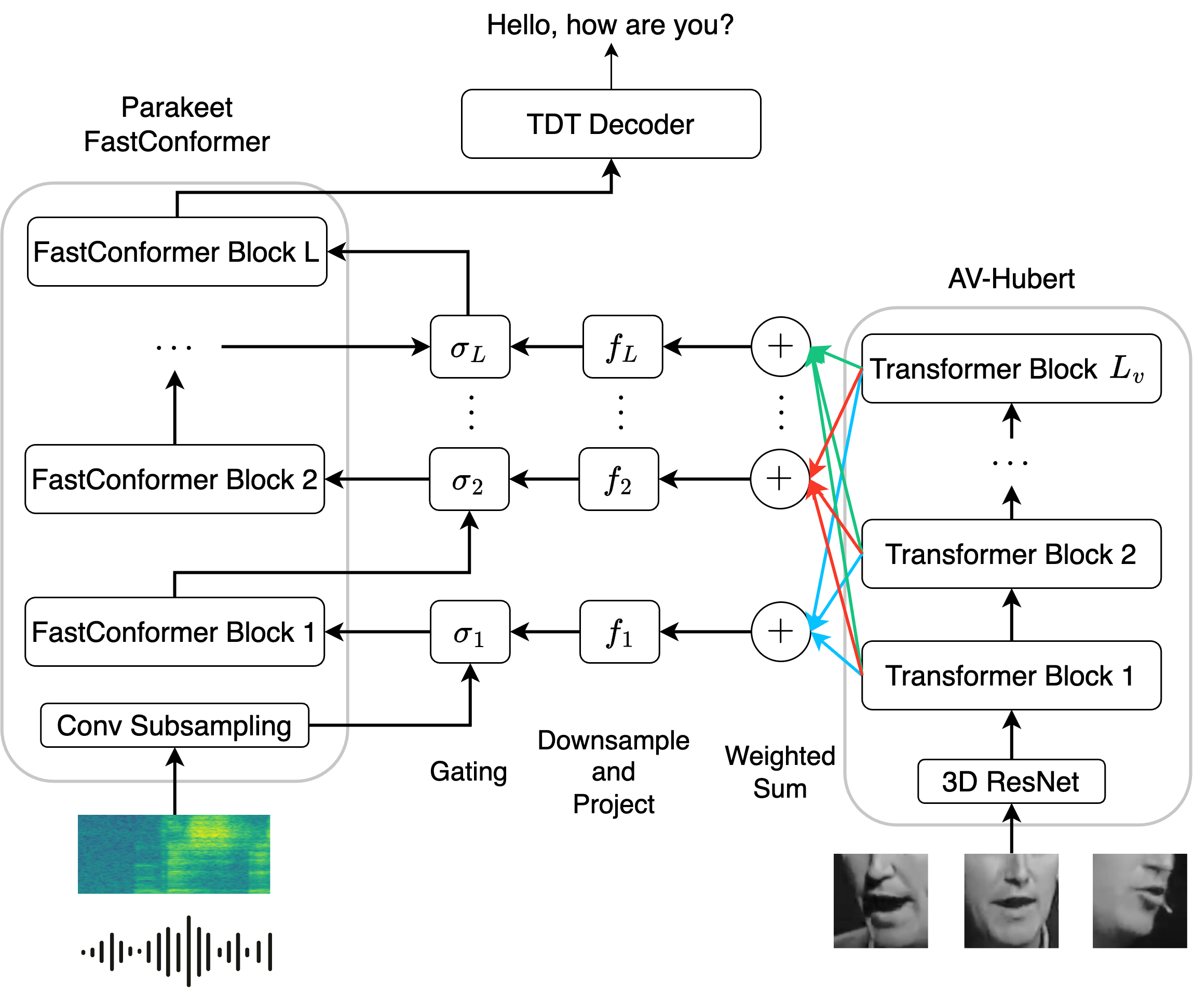}
    \caption{Schema of the proposed AV-TS-ASR system describing the visual conditioning. Different arrow colors represent different sets of learned weights that are used in AV-Hubert hidden representation averaging in the Equation~\ref{eq:weighted_avg}. $f_l$ blocks correspond to the Equation~\ref{eq:vis_processing} and $\sigma_l$ block correspond to the Equation~\ref{eq:gating}.}
    \label{fig:system_schema}
\end{figure}

The proposed AV-TS-ASR architecture is shown in Figure~\ref{fig:system_schema}.
Given a multi-talker audio mixture $a \in \mathbb{R}^{T_a}$ and a video stream $v_s \in \mathbb{R}^{T_v \times H \times W}$ corresponding to a target speaker $s$, 
we extract representations from the $l$-th layer of the  Parakeet-tdt-0.6b-v2 FastConformer encoder~\cite{rekesh2023fastconformer} and the $k$-th layer of the AV-Hubert visual encoder:
\begin{equation}
    e_a^l = \text{PKT}(a, l) \in \mathbb{R}^{N_a \times d_a},
\end{equation}
\begin{equation}
    e_{v_s}^k = \text{AVH}(\mathbf{0}, v_s, k) \in \mathbb{R}^{N_v \times d_v},
\end{equation}
where $N_a$ and $N_v$ represent the sequence lengths of the acoustic and visual features, and $d_a, d_v$ are their dimensions. We zero out the audio input to AV-HuBERT to utilize it strictly as a visual encoder for the target speaker $s$.

To enable the model to select the visual information most relevant to each depth of the acoustic encoder, we first compute a weighted sum of all the AV-HuBERT layers for each acoustic layer $l$ using non-negative learnable weights $\alpha_{i}^l$ that sum to 1:

\begin{equation}
    \label{eq:weighted_avg}
    f_{v_s}^l = \sum_{i}^{L_v} \alpha_{i}^l e_{v,s}^i.
\end{equation}

Subsequently, to align these aggregated features with the acoustic representations in terms of temporal rate (12.5\,Hz vs. 25\,Hz) and embedding space, we apply a 1D convolution (kernel size 5, stride 2), followed by a feed-forward network (FFN)~\cite{vaswani2017attention} with a SiLU activation~\cite{elfwing2018silu}, dropout~\cite{srivastava2014dropout} ($p=0.1$), and layer normalization~\cite{ba2016layernormalization} at the end:
\begin{equation}
    \label{eq:vis_processing}
    \tilde{f}_{v_s}^l = \text{FFN}(\text{Conv1D}(f_{v_s}^l)) \in \mathbb{R}^{N_a \times d_a}.
\end{equation}

We then fuse the aligned modalities using a multiplicative gating mechanism. We concatenate the normalized acoustic features and the aligned target visual features to compute a gate $g^l \in (0, 1)$:
\begin{equation}
    \label{eq:gating}
    g^l = \sigma(W_{gate} [\text{LN}(e_a^l) \parallel \tilde{f}_{v_s}^l] + b_{gate}) \in \mathbb{R}^{N_a \times d_a}.
\end{equation}

The fused representation, which serves as the input to the next layer, is a gated combination defined as:
\begin{equation}
    e_{out}^l = g^l \odot e_a^l + (1 - g^l) \odot \tilde{f}_{v_s}^l \in \mathbb{R}^{N_a \times d_a},
\end{equation}
where $\odot$ denotes element-wise multiplication. This gate dynamically controls the reliance on the target speaker's lip-reading cues versus the mixed audio at every frame. 

Finally, the sequence of contextualized representations $e_{out}^L$ from the last encoder layer is passed to the Time-and-Duration Transducer (TDT)~\cite{xu2023tdt} decoder. During training, we optimize the following negative log-likelihood:
\begin{equation}
    \mathcal{L} = -\log P(\y_s \mid a, v_s),
\end{equation}
where $\y_s$ is the ground-truth target-speaker transcript.

\subsection{Conversation Group Clustering}
\label{sec:conv_group_clustering}

While the challenge baseline clustering algorithm relies solely on speech overlap duration—assuming that frequent overlaps among speakers indicate different conversation groups—this heuristic fails to account for semantic context and often misclassifies non-overlapping speakers who belong to the same conversation. To address this, we propose a multi-stage clustering pipeline that leverages the semantic capabilities of Qwen3.5 using the transcripts generated by our AV-TS-ASR system. The pipeline proceeds in three stages: distinguish active participants from passive listeners, cluster active participants via semantic similarity, and assign passive listeners to groups using a speech overlap-based fallback.

\subsubsection{Topic Detection and Semantic Clustering}
\label{subsec:topic_clustering}

First, we distinguish between speakers that are active and those that are passive and only produce backchannel cues (e.g., "yeah", "mhm"). For each speaker $s$, we input their full transcript $\y_s$ to the LLM and prompt it to output hard binary classification decisions: \textit{True} if the speaker speaks about some topic, and \textit{False} otherwise.

For every pair of active speakers $(s_i, s_j) \in S_{active}$, we prompt the LLM to estimate a topic similarity score $\sigma_{i,j} \in [0, 1]$, where 1 indicates identical topics. We construct a similarity matrix from these scores and apply Agglomerative Hierarchical Clustering (AHC) with a threshold of $\tau=0.7$ estimated on the MCoRec training set. This process yields a set of core conversation clusters $\mathcal{C} = \{C_1, C_2, \dots, C_K\}$.

\subsubsection{Time-based Fallback}
\label{subsec:time_based_fallback}

For speakers who lack sufficient semantic content, we utilize speech overlap similarly to the MCoRec baseline to assign the passive speakers to the core clusters $\mathcal{C}$. We represent each cluster $C_k$ as a single ``pseudo-speaker'' and construct a distance matrix for the combined set of passive speakers and pseudo-speakers.

To retain the original clusters in $C$, the distance between any two pseudo-speakers is fixed at $1.0$ (maximum distance). The distance between a passive speaker $s_p$ and a cluster $C_k$ is defined as the average overlap ratio between $s_p$ and all the speakers in $C_k$:
\begin{equation}
    d(s_p, C_k) = \frac{1}{|C_k|} \sum_{s_q \in C_k} \frac{\text{overlap}(s_p, s_q)}{\text{duration}(s_p, s_q)},
\end{equation}
where $\text{overlap}(s_p, s_q)$ is the duration of simultaneous speech and $\text{duration}(s_p, s_q)$ is the total speech time. Applying AHC to this distance matrix assigns passive speakers to the conversation groups with which they have the least speech overlap and ensures that active speakers are clustered according to topic similarity.

\section{Experimental Setup}
\label{sec:exp_setup}

We developed the AV-TS-ASR system using the NeMo toolkit~\cite{kuchaiev1909nemo}. For the LLM-based clustering pipeline, we utilized the DSPy framework~\cite{khattab2024dspy} to systematically structure prompts and modularize the workflow.

\subsection{Data Simulation}
The proposed system is pre-trained in two stages to progressively adapt the newly-introduced parameters.

For the first pre-training stage, we generated LibriMix-style~\cite{cosentino2020librimix} fully overlapped mixtures using the AVYT~\cite{nguyen25b_interspeech} and LRS3~\cite{afouras2018lrs3} datasets. We discarded segments falling outside the 5th percentile of duration and applied bucket sampling with 10 uniform buckets to maximize speech overlap. The dataset comprises 200k mixtures with 2--3 speakers and 100k mixtures with 4 speakers. Additionally, we augmented these with silent AVYT segments to train the model to perform visual speaker activity detection. This resulted in approximately 1500\,h of pre-training data.

For the second pre-training stage, we used AMI~\cite{carletta2005ami} single distant microphone (SDM) audio-visual recordings to approximate the MCoRec conditions where multiple distinct conversations occur simultaneously. We randomly overlapped 2--4 disjoint sessions to create ``cocktail party'' scenarios with independent conversational groups, yielding approximately 150\,h of data. For both stages, we constructed development sets from the held-out data to facilitate checkpoint selection.

\subsection{MCoRec Data Preprocessing}
During inspection of the MCoRec dataset, we identified several mismatched reference transcripts. In the development set, the transcripts for $(54,4)$ and $(55,3)$ session-speaker pairs were incorrectly replaced by those of $(53,2)$ and $(54,4)$, respectively. As we could not recover the correct transcript for $(53,2)$, we removed that example. In the training set, we identified additional mismatches in $(26,2)$, $(27,0)$, $(28,2)$, $(29,2)$, and $(30,2)$, which we removed to avoid label noise. The corrected data were used only for training and model selection, while all results in Section~\ref{sec:results} are reported on the original MCoRec data to make the future work comparisons easier.

\subsection{Augmentations}
For visual inputs, we randomly cropped $96\times 96$ lip videos to $88\times88$ pixels and applied Gaussian blur to simulate focus variations. To account for artifacts from face crop preprocessing, we perturbed each frame with random rotations and brightness changes modeled by a mean-reverting stochastic process. Furthermore, we employed random span-based masking on both audio and video inputs to prevent the model from over-relying on a single modality.

\subsection{Training Curriculum}
First, we fine-tuned AV-HuBERT on simulated visual-only data with a frozen pre-trained Parakeet-v2 RNN-T TDT decoding head to improve visual modeling. We then adopted a two-phase strategy to introduce visual conditioning without degrading the pre-trained acoustic representations; AV-HuBERT and the RNN-T joiner and predictor were kept frozen throughout these stages.

In the first stage, we trained the newly introduced parameters on the 1500\,h of simulated overlap dataset while freezing both the Parakeet and AV-HuBERT encoders. This prevented the backpropagation of ill-posed gradients from randomly-initialized parameters to the pre-trained acoustic encoder. After convergence, we continued pre-training on the same data with the acoustic encoder unfrozen.

In the second stage, we resumed from the best first-stage checkpoint and trained on a concatenation of simulated AMI data and the MCoRec training set. Preliminary experiments showed that this outperformed sequential fine-tuning on AMI followed by MCoRec.

\subsection{Training Details}

We trained all models in bfloat16 on NVIDIA A100 and H100 GPUs using the AdamW optimizer~\cite{loshchilov2018adamw} with the Noam learning-rate (LR) scheduler~\cite{vaswani2017attention}, a weight decay of $10^{-2}$, and 10k warm-up steps. During the first pre-training stage, the peak LR for newly introduced parameters was set to $2.5 \times 10^{-5}$, while the Parakeet FastConformer parameters used a learning rate five times smaller. For fine-tuning, the learning rate was reduced by a factor of two relative to pre-training.

Utterance length was not constrained during the first pre-training stage as the maximum utterance length was approximately 20\,s. For fine-tuning, we sampled random 50-second segments to expose the model to higher intra-utterance variability. We experimented with longer segments but observed no improvement on the development set.

\subsection{Inference}
First, we chunk the input video into 20\,s long non-overlapping segments to infer the AV-HuBERT visual features, as the model does not support long-form inference. We selected the chunk size based on the development set performance.

Afterwards, the visual features are concatenated along the time dimension and are fused into the Parakeet FastConformer acoustic features. 

Lastly, RNN-T TDT decodes the entire target-speaker feature sequence using greedy decoding.

\subsection{Prompting}
We use the DSPy framework, which allows us to specify the input, output, and task description via a DSPy Signature .\footnote{\href{https://dspy.ai/learn/programming/signatures/}{https://dspy.ai/learn/programming/signatures/}} 
The framework then handles prompt construction, LLM API invocation, and output parsing. To improve reproducibility and reduce hallucinations, we set the decoding temperature to 0.

\subsubsection{Topic Detection Task}
To detect if the speaker is active, the LLM is provided the following field described as: \texttt{transcript} - \textit{a single speaker transcript}. The model is then required to predict \texttt{contains\_topic} described as: \textit{Whether it is possible to detect topic or subject from the transcript}.

\subsubsection{Pairwise Topic Similarity Task}
To estimate the topic similarity score, the LLM takes one input variable \texttt{transcripts} described as: "A dictionary mapping speaker IDs to their transcripts. Each key is a speaker ID.". Then the model predicts \texttt{topic\_similarity} described as: "Score between 0-1 indicating topic similarity between the two speakers. 0 = completely different topics, 1 = same topic.". 

We observed during preliminary experiments that the model adheres to the specified ranges and always outputs a topic similarity score between 0 and 1; hence, no post-processing was used.

\section{Results}
\label{sec:results}

    We submitted two systems of the same architecture, both pre-trained on the 1500\,h simulated dataset but differing in the fine-tuning data. \textbf{System~1} was fine-tuned on the MCoRec training set and simulated AMI mixtures, while \textbf{System~2} was fine-tuned on the MCoRec training set only. Both systems used the same clustering.

\begin{table}[hbt]
    \centering
    \caption{Performance comparison on the MCoRec development set. The Joint Score combines WER and clustering performance as defined by the challenge metric.}
    \label{tab:system_comparison}
    \begin{tabular}{lccc}
        \toprule
        System & WER~$\downarrow$ & F1~$\uparrow$ & Joint Score~$\downarrow$ \\
        \midrule
        Baseline & 49.90\% & 0.815 & 0.355 \\
        \midrule
        System 1 & 33.87\% & 0.967 & 0.185 \\
        System 2 & 33.69\% & 0.967 & 0.184 \\
        \bottomrule
    \end{tabular}
\end{table}

Table~\ref{tab:system_comparison} presents the main results. Both proposed systems significantly outperform the challenge baseline in transcription and clustering accuracy. Interestingly, despite seeing approximately $10\times$ more fine-tuning data, System~1 performs slightly worse than System~2. We attribute this to the domain mismatch between the simulated AMI mixtures and the target MCoRec data, as well as the lower video quality of the AMI corpus, which was recorded over two decades ago. Consequently, we selected System~2 for subsequent ablation studies.

\begin{table}[bt]
    \centering
    \caption{Comparison of inference strategies on the MCoRec development set using System 2.}
    \label{tab:inference_strategies}
    \begin{tabular}{lc}
        \toprule
        Inference Mode & WER~$\downarrow$ \\
        \midrule
        Full Long-form & 33.69\% \\
        Per track & 33.93\% \\
        According to ASD & 67.93\% \\
        \bottomrule
    \end{tabular}
\end{table}

Table~\ref{tab:inference_strategies} analyzes the impact of inference strategy. The meeting videos were split to tracks according to the missing face detection frames. ``Full Long-form'' concatenates tracks into a continuous stream and fills the gaps using black frames, while ``Per Track'' decodes each track independently. The slight improvement of long-form inference suggests that the model leverages acoustic context even when visual information is not available. In contrast, segmentation based on the baseline Active Speaker Detection (ASD) yields poor results. This degradation occurs because ASD segmentation often produces short chunks ($<5$\,s), creating a mismatch, as our model was trained on $\approx 50$\,s segments.

\vspace{-10pt}
\begin{table}[hbt]
    \centering
    \caption{Ablation study on the effect of pre-training data composition.}    \label{tab:amnt_of_pretraining_data}
    \begin{tabular}{lc}
        \toprule
        Training Data & WER~$\downarrow$ \\
        \midrule
        Simulated pre-training + MCoRec & 33.69\% \\
        Simulated pre-training + AMI mixtures & 37.87\% \\
        AMI mixtures + MCoRec & 39.85\% \\
        MCoRec & 38.20\% \\
        \bottomrule
    \end{tabular}
\end{table}

Table~\ref{tab:amnt_of_pretraining_data} highlights the critical role of synthetic data. Pre-training on 1500\,h of simulated mixtures provides a substantial boost compared to training on MCoRec alone. Notably, a model trained solely on simulated data (Simulated pre-training + AMI Mixtures) achieves a WER of 37.87\% without ever seeing the real MCoRec data. This zero-shot generalization (only 4.18\% absolute WER degradation compared to the fine-tuned model) confirms the efficacy of our simulation pipeline. Finally, removing the pre-training stage and training on around 155\,h of data from AMI and MCoRec (Row 3) results in a higher WER compared to training with simulated pre-training, demonstrating that the 1500\,h of simulated data significantly improves the performance. Similarly to the system comparison in Table~\ref{tab:system_comparison}, we observe that training on MCoRec alone results in 1.65\% absolute WER improvement, likely due to the domain mismatch between AMI and MCoRec.

\begin{table}[hbt]
    \centering
    \caption{Conversation clustering F1 Scores (\%) on MCoRec using ground-truth transcripts. $\text{Baseline}^*$ denotes clustering applied to time spans derived from ground-truth transcripts rather than ASD output.}    \label{tab:clustering_comparison}
    \begin{tabular}{lll}
        \toprule
        System      & Train F1~$\uparrow$ & Dev F1~$\uparrow$ \\
        \midrule
        Baseline   & 85.87    & 81.53  \\
        $\text{Baseline}^*$    & 73.75    & 79.14  \\
        \midrule
        Joint       & 91.13    & 86.04  \\
        Pairwise    & 97.94    & 98.46  \\
        Pairwise With Fallback & 97.94    & 98.46 \\ 
        \bottomrule
    \end{tabular}
\end{table}

Table~\ref{tab:clustering_comparison} reports clustering performance under different setups. To isolate clustering effects from recognition errors, we use ground-truth transcripts.

The first two rows indicate that ASD-derived time spans are more suitable for clustering than spans obtained from the reference transcripts. The third row presents a naive LLM-based approach where the model is tasked to clusters all the speakers based on their transcripts, which already provides a substantial improvement over the baseline. The fourth row performs pairwise topic-based clustering and further improves performance over the joint formulation, suggesting benefits from simplifying the decision to pairwise similarity estimation.

The final configuration augments the pairwise topic-based method with an overlap-based fallback. The results imply that passive speakers are not present in the training or development data. Nevertheless, we retain this component in the final submission to avoid assuming the same condition for the evaluation set.

\begin{figure}[t]
    \centering
    \includegraphics[width=0.8\linewidth]{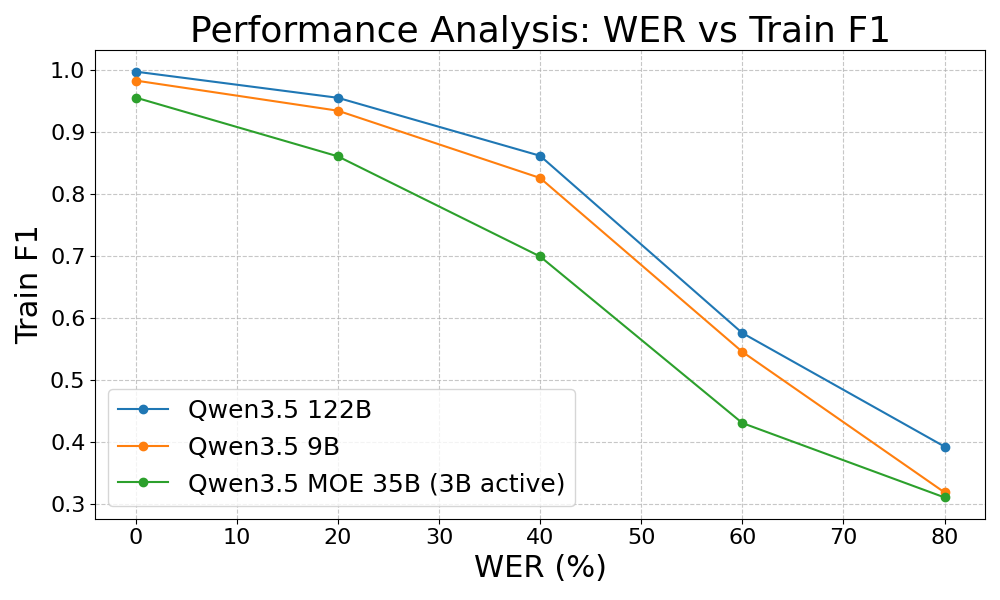}
    \vspace{-10pt}
    \caption{F1 score of pairwise topic scores inferred using different QWEN3.5 model sizes against transcription errors.}\
    \vspace{-15pt}
    \label{fig:wer_f1_comparison}
\end{figure}

Lastly, we simulated random transcription errors to analyze the clustering robustness. Figure~\ref{fig:wer_f1_comparison} compares how different LLM sizes handle transcription errors. As expected, the 122B model performs the best compared to the smaller models and suggest that WER up to around 30\% still results in F1 score above 90\%. However, the performance degradation is not proportional to the number of active parameters, suggesting that smaller LLMs can perform the topic similarity score estimation well.

\section{Conclusion}
\label{ssec:subhead}
We presented a long-form Audio-Visual Target-Speaker ASR system for the CHiME-9 MCoRec task, leveraging the strong pre-trained representations of NVIDIA Parakeet-v2 and AV-HuBERT. By fusing these modalities via a gated adapter and replacing heuristic clustering with an LLM-driven semantic approach, we achieved substantial improvements over the official challenge baseline. Future work will focus on improving the quality of simulated data and exploring alternative visual backbones to further close the domain gap.

\vfill\pagebreak

\bibliographystyle{IEEEbib}
\bibliography{strings,refs}

\end{document}